\documentclass[12pt,preprint,useAMS,usenatbibi,usegraphicx]{emulateapj}
\usepackage{latexsym}
\usepackage{amssymb}
\usepackage{amsmath}

\def\msun{\,{\rm M_\odot}}

\def\etal{{et al.\ }}

\newcommand\beq{\begin{equation}}
\newcommand\eeq{\end{equation}}
\newcommand{\ba}{\begin{eqnarray}}
\newcommand{\ea}{\end{eqnarray}}  
\def\spose#1{\hbox to 0pt{#1\hss}}
\def\lta{\mathrel{\spose{\lower 3pt\hbox{$\mathchar"218$}}
     \raise 2.0pt\hbox{$\mathchar"13C$}}}
\def\gta{\mathrel{\spose{\lower 3pt\hbox{$\mathchar"218$}}
 \raise 2.0pt\hbox{$\mathchar"13E$}}}

\shorttitle{Massive black hole binaries}
\shortauthors{Sesana, Haardt, \& Madau}
\slugcomment{ApJ, in press}

\begin{document}

\title{Interaction of massive black hole binaries with their stellar environment: II. 
Loss-cone depletion and binary orbital decay}
\author{Alberto Sesana\altaffilmark{1}, Francesco Haardt\altaffilmark{1}, \&
Piero Madau\altaffilmark{2,3}}

\altaffiltext{1}{Dipartimento di Fisica e Matematica, Universit\'a dell'Insubria, via 
Valleggio  11, 22100 Como, Italy.}
\altaffiltext{2}{Department of Astronomy \& Astrophysics, University of 
California, 1156 High Street, Santa Cruz, CA 95064.}
\altaffiltext{3}{Max-Planck-Institut fuer Astrophysik, 
Karl-Schwarzschild-Strasse 1, 85740 Garching bei Muenchen, Germany.}

\begin{abstract}
We study the long-term evolution of massive black hole binaries (MBHBs) 
at the centers of galaxies using detailed scattering experiments to solve the full three-body problem.
Ambient stars drawn from a isotropic Maxwellian distribution unbound to the binary 
are ejected by the gravitational slingshot. We construct a minimal, hybrid 
model for the depletion of the loss cone and the  
orbital decay of the binary, and show that secondary slingshots --
stars returning on small impact parameter orbits 
to have a second super-elastic scattering with the MBHB -- may considerably 
help the shrinking of the pair in the case of large binary mass ratios.  
In the absence of loss-cone refilling by two-body
relaxation or other processes, the mass ejected before the stalling of a 
MBHB is half the binary reduced mass. About 50\% of the ejected stars are expelled 
in a ``burst" lasting $\sim 10^4\,{\rm yr}\,M_6^{1/4}$, where $M_6$ 
is the binary mass in units of $10^6 \msun$. The loss cone is completely emptied 
in a few bulge crossing timescales, $\sim 10^7\,{\rm yr}\,M_6^{1/4}$. 
Even in the absence of two-body relaxation or gas dynamical processes, 
unequal mass and/or eccentric binaries with $M_6\gtrsim 0.1$ can shrink 
to the gravitational wave emission regime in less than a Hubble time, 
and are therefore ``safe" targets for the planned {\it Laser Interferometer 
Space Antenna (LISA)}.  
\end{abstract}


\section{Introduction}

It is now widely accepted that the formation and evolution of galaxies and
massive black holes (MBHs) are strongly linked: MBHs are ubiquitous in the 
nuclei of nearby galaxies, and a tight correlation is observed between hole 
mass and the stellar mass of the surrounding spheroid or bulge (e.g. Magorrian 
\etal 1998; Gebhardt et al. 2000; Ferrarese \& Merritt 2000; Haring \& Rix 2004). 
If MBHs were also common in the past (as implied by the notion that distant 
galaxies harbor active nuclei for a short period of their life), and if their host
galaxies experience multiple mergers during their lifetime, as dictated by
cold dark matter (CDM) hierarchical cosmologies, then close MBH binaries (MBHBs) 
will inevitably form in large numbers during cosmic history (Begelman, Blandford,
\& Rees 1980). Observations 
with the {\it Chandra} satellite have indeed revealed two active MBHs in the nucleus of 
NGC 6240 (Komossa et al. 2003), and a MBHB is inferred in the radio core 
of 3C 66B (Sudou et al. 2003). The VLBA discovery in the radio galaxy 0402+379 of 
a MBHB system with a projected separation of just 7.3 pc has recently been reported
by Rodriguez \etal (2006). 
MBH pairs that are able to coalesce in less than a Hubble time will 
give origin to the loudest gravitational wave (GW) events in the universe.
In particular, a low-frequency space interferometer like the planned {\it Laser 
Interferometer Space Antenna (LISA)} is expected to have the sensitivity to 
detect nearly all MBHBs in the mass range $10^4-10^7\,\msun$ that happen to 
merge at any redshift during the mission operation phase (Sesana et al. 
2005). The coalescence rate of such ``{\it LISA} MBHBs'' depends, however, 
on the efficiency with which stellar and gas dynamical processes can drive 
wide pairs to the GW emission stage. 

Following the merger of two halo$+$MBH systems of comparable 
mass (``major mergers''), it is understood that dynamical friction will drag 
in the satellite halo (and its MBH) toward the center of the more massive 
progenitor (see, e.g., Kazantzidis et al. 2005): this will lead to the 
formation of a bound MBH binary in the violently relaxed core of the newly 
merged stellar system. As the binary separation decays, the effectiveness of 
dynamical friction slowly declines because distant stars perturb the 
binary's center of mass but not its semi-major axis (Begelman \etal 1980). 
The bound pair then hardens by capturing stars passing  in its immediate 
vicinity and ejecting them at much higher velocities (gravitational slingshot).
It is this phase that is considered the bottleneck of a MBHB's path to 
coalescence, as there is a finite supply of stars on intersecting orbits
and the binary may ``hung up'' before the back-reaction from GW emission becomes
important. This has become known as the ``final parsec problem'' 
(Milosavljevic \& Merritt 2003, hereafter MM03).

While the final approach to coalescence of binary MBHs is still not well 
understood, several computational tools have been developed to tackle the 
problem at hand. The orbital decay rate depends on several parameters of the 
guest binary (mass, mass ratio, orbital separation, and eccentricity), and 
on the stellar distribution function of the host galaxy bulge. In early treatments 
(e.g. Mikkola \& Valtonen 1992; Quinlan 1996, hereafter Q96),
the stellar ejection rate and the rate of change of the binary semi-major axis 
and eccentricity were derived via three-body scattering experiments in a 
fixed stellar background. The assumption of a fixed background breaks down 
once the binary has ejected most of the stars on intersecting orbits, and 
the extraction of energy and angular momentum from the binary can continue 
only if new stars can diffuse into low-angular momentum orbits (refilling 
the binary's phase-space ``loss cone''), or via gas processes (Escala et 
al. 2004; Dotti, Colpi, \& Haardt 2006). Hybrid approaches in which the rate 
coefficients derived from numerical experiments in a fixed background are 
coupled with a model for loss-cone repopulation have been used, e.g., 
by Yu (2002) and MM03, while the limiting case in which the loss cone 
is constantly refilled but the central stellar density decreases due to 
mass ejection has been studied in a cosmological context by Volonteri, 
Haardt, \& Madau (2003) and Volonteri, Madau, \& Haardt (2003). 
A fully self-consistent, N-body approach to the evolution of MBHBs, while 
clearly desirable, is limited today to $N\lesssim 10^6$ particles,  
corresponding to a mass resolution of $m_*/M\sim 10^{-3}$ 
(e.g. Quinlan \& Hernquist 1997; Milosavljevic \& Merritt 2001; Hemsendorf, 
Sigurdsson, \& Spurzem 2002; Aarseth 2003; Chatterjee, Hernquist, \& Loeb 
2003; Makino \& Funato 2004; Berczik, Merritt, \& Spurzem 2005). Such 
performance figures are not sufficient to reproduce central bulges, even 
of faint galaxies; and the small particle numbers cause an artificial 
enhancement of star-star scatterings and of the Brownian motion of the 
binary, leading to a spurious refilling of the loss cone. 

This is the second paper in a series aimed at a detailed study of the
interaction of MBHBs with their stellar environment. 
In Sesana, Haardt \& Madau (2006, hereafter Paper I), three-body scattering 
experiments were performed to study the ejection of hypervelocity stars (HVSs) by MBHBs 
in a fixed stellar background. In this paper, we use a hybrid approach 
to investigate the orbital decay and shrinking of MBHBs in time-evolving stellar cusps. 
Numerically derived rates of stellar ejections stars are coupled to an extension of 
the analytical formulation of loss-cone dynamics given by MM03. This method 
allows us to simultaneously follow the orbital decay of the pair as well 
as the time evolution of the stellar distribution function. 
MBHBs are embedded in the deep potential wells of 
galaxy bulges, so when the binary first becomes ``hard'' only a few stars 
acquire a kick velocity large enough to escape the host. The bulge 
behaves as a collisionless system, and many ejected stars will return to the 
central region on nearly unperturbed, small impact parameter orbits, and 
will undergo a second super-elastic scattering with the binary, as first discussed by MM03. 
Under the assumption of a spherical potential, we 
quantify the role of these ``{\it secondary slingshots}'' in determining the
hardening of the pair. The plan of the paper is as 
follows. In \S~2 we describe our hybrid model for the orbital evolution of 
MBHBs in a time evolving stellar density profile. The shrinking and coalescence 
of the binary is discussed in \S~3. 

\section{Hardening in a time-evolving background \label{sc:tis}}
\subsection{Scattering experiments}

Our hybrid method relies on the large number of outputs from the suite of three-body 
scattering experiments presented in Paper I. In the following, we briefly summarize 
the basic theory. Consider a binary of mass $M=M_1+M_2=M_1(1+q)$ ($M_2\leq M_1$), 
reduced mass 
$\mu=M_1M_2/M$, and semimajor axis $a$, orbiting in a background of stars of mass
$m_*$. In the case of a light intruder with $m_* \ll M_2$,
the problem is greatly simplified by setting the center of mass of
the binary at rest at the origin of the coordinate system.
It is then convenient to define an approximate
dimensionless energy change $C$ and angular momentum change $B$ in a single
binary-star interaction as (Hills 1983)
\begin{equation}\label{c}
C=\frac{M}{2m_*}\frac{\Delta E}{E}=\frac{a\Delta E_*}{G\mu},
\end{equation}
and
\begin{equation}\label{b}
B=-\frac{M}{m_*}\frac{\Delta L_z}{L_z}=\frac{M}{\mu}\frac{\Delta L_{z*}}{L_z}.
\end{equation}
Here $\Delta E/E$ is the fractional increase (decrease if negative)
in the orbital specific binding energy $E=-GM/(2a)$, $\Delta L_z/L_z$
is the fractional change in orbital specific angular momentum 
$L_z=\sqrt{GMa(1-e^2)}$, while
$\Delta E_*$ and $\Delta L_{z*}$ are the corresponding
changes for the interacting star. The quantities $B$ and $C$ are of order 
unity and can be derived by three-body scattering experiments that 
treat the star-binary encounters one at a time (Hut \& Bahcall 1983; Q96). 
For each encounter one solves nine coupled, second-order, differential equations 
supplied by 18 initial conditions. The initial conditions define a point in a 
nine-dimensional parameter space represented by the mass ratio $q=M_2/M_1$ of the 
binary, its eccentricity $e$, the mass of the incoming field star, its
asymptotic initial speed $v$, its impact parameter at infinity $b$, and 
four angles describing the initial direction of the impact, its initial 
orientation, and the initial binary phase. 
A significant star-binary energy exchange (i.e. characterized by a dimensionless
energy change $C>1$) occurs only for $v<V_c\sqrt{M_2/M}$, where $V_c=\sqrt{GM/a}$ 
is the binary orbital velocity
(the relative velocity of the two holes if the binary is circular, see e.g. Saslaw, 
Valtonen, \& Aarseth 1974; Mikkola \& Valtonen 1992). 

A set of 24 scattering experiments was performed for different binary mass ratios
and initial eccentricities, each run tracking the orbital evolution of 
$4\times 10^6$ stars. The binary evolution in an isotropic stellar background
of density $\rho$ and one-dimensional velocity dispersion $\sigma$ at infinity
is determined by three dimensionless quantities (Q96): the hardening rate
\begin{equation}\label{dadtmaxw}
H={\sigma\over G\rho}\frac{d}{dt}\left(\frac{1}{a}\right),
\end{equation}
the mass ejection rate ($M_{\rm ej}$ is the stellar mass ejected by the
binary)
\begin{equation}\label{jsigma}
J=\frac{1}{M} \frac{dM_{\rm ej}} {d\ln(1/a)},
\end{equation}
and the eccentricity growth rate
\begin{equation}\label{de}
K=\frac{de}{d\ln(1/a)}.
\end{equation}
The hardening rate $H$ is approximatively constant for separations smaller than
the ``hardening radius",  
\begin{equation}\label{hardrad}
a<a_h=\frac{GM_2}{4\sigma^2}
\end{equation}
(Q96). The binary is assumed to be embedded in
a bulge of mass $M_B$, radius $R_B$, and stellar density profile
approximated by a singular isothermal sphere (SIS).
Stars are counted as ``ejected'' from the bulge if, after three-body
scattering, their velocity $V$ far away from the binary is greater
than the escape velocity from the radius of influence of the binary,
$r_{\rm inf}\equiv GM/(2\sigma^2)$. The
SIS potential is $\phi(r)=-2\sigma^2\,[\ln (GM_B/2\sigma^2r)+1]$
(for $r<R_B=GM_B/2\sigma^2$), and the escape speed from $r_{\rm inf}$ is then
\begin{equation}
\begin{split}
v_{\rm esc}&\equiv \sqrt{-2\phi(r_{\rm inf})}\\
&=2\sigma \sqrt{[\ln(M_B/M)+1]}=5.5\sigma,
\end{split}
\label{eq:vesc}
\end{equation}
where the second equality comes from the adopted bulge-black hole mass relation
$M=0.0014\,M_B$ (Haring \& Rix 2004). Stars that do not acquire a kick velocity 
large enough to escape the host bulge, i.e. with
\begin{equation}
\begin{split}
V<v_{\rm ret}&=\sqrt{2\phi(R_B)-2\phi(r_{\rm inf})}\\
&=2\sigma\sqrt{\ln(M_B/M)}=\,5.1\sigma,
\end{split}
\label{eq:vret}
\end{equation} 
are allowed multiple interactions with the binary. They can return to the central regions 
on small impact parameter orbits and undergo a second super-elastic scattering
(``secondary slingshots''). Secondary scatterings are not allowed for stars ejected with 
$5.1\,\sigma \lta V \lta 5.5\,\sigma$, since even a small deviation from sphericity of the 
galaxy gravitational potential would make them miss the shrinking MBHB on their 
return to the center. Note that, even if they were able to undergo another interaction 
with the binary, such stars would not contribute significantly to binary hardening as long as 
the condition $V>V_c\sqrt{M_2/M}=2\sigma\sqrt{a_h/a}$ is satisfied, since in this case
the star-binary energy exchange would be negligible. 

\begin{figure}
\epsscale{1.0}
\plotone{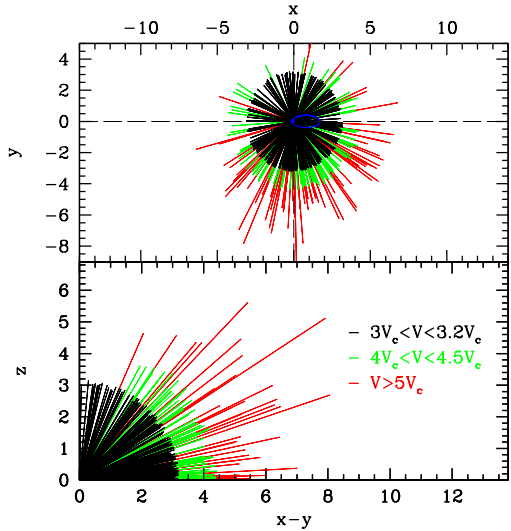}
\caption{Velocity diagram of scattered stars in three different speed ranges: 
$3V_c<V<3.2V_c$ ({\it black vectors}), $4V_c<V<4.5 V_c$ ({\it green vectors}), 
and $V>5V_c$ ({\it red vectors}). {\it Upper panel}: longitude diagram of scattered stars. Each
vector length is proportional to the modulus of the star's total velocity (not to the 
velocity projected into the $xy$ plane). The blue ellipse shows the counter-clockwise 
orbit of the lighter black hole of the binary.
{\it Lower panel}: latitude diagram of scattered stars. 
}
\label{fig:ejected}
\end{figure}

As shown in Figure \ref{fig:ejected} and discussed in details in Paper I, 
three-body interactions create a
subpopulation of HVSs on nearly radial orbits, with a spatial distribution
that is initially highly flattened in the inspiral plane of the binary, but
becomes more isotropic with decreasing binary separation. The degree of
anisotropy is smaller for unequal mass binaries and larger for stars with
higher kick velocities. Eccentric MBHBs produce a more prominent tail of
high-velocity stars and break axisymmetry, ejecting HVSs along a
broad jet perpendicular to the semimajor axis. The jet two-sidedness
decreases with increasing binary mass ratio, while the jet opening-angle
increases with decreasing kick velocity and orbital separation.

\subsection{Loss-cone time evolution}

In the absence of loss-cone refilling by two-body relaxation or other
processes, the supply of stars that can interact with the black hole pair 
is limited. Analytic expressions for non-equilibrium loss-cone dynamics
based on the evolution of the stellar distribution function as a result 
of repeated ejections have been given in MM03. Here we adopt a hybrid 
approach instead, combining the results of scattering experiments with an extension 
of MM03's study. 

\subsubsection{Stellar content}

The stellar content of the loss cone can be estimated from simple geometrical 
considerations. When the MBHB separation is $a\lta a_h$, only a small fraction of bulge stars 
have low-angular momentum trajectories with pericenter distance $r_p<a$.
In a spherical velocity distribution, the 
fraction of trajectories originating at $r$ and crossing a sphere of radius 
$a<r$ around the center is
\begin{equation}\label{theta}
\Theta(r)=\left[1-\sqrt{1-\left(\frac{a}{r}\right)^2}\right].
\end{equation}
The stellar mass within the geometrical loss cone is then
\begin{equation}\label{lossconeint}
M_*=\int_0^{a}4\pi r^2\,\rho(r)\,dr+\int_{a}^{R_B}4\pi r^2\,
\rho(r)\Theta(r) \,dr.
\end{equation}
For a SIS $\rho(r)=\sigma^2/(2\pi Gr^2)$, and equation (\ref{lossconeint}) is readily 
integrated to yield in the limit $a\ll R_B$
\begin{equation}\label{lossconeSIS}
M_*\simeq{\pi\sigma^2\over G}\,a=\frac{\pi}{4}\left(\frac{a}
{a_h}\right)\,M_2.
\end{equation}

The above scheme is oversimplified, as it assumes 
stellar trajectories to be straight lines. The gravitational field of the 
stellar mass distribution increases the net number of distant stars with 
pericenter distances $r_p<a$. Consider a star at distance  $r>a$ from the binary
moving with random velocity $v$. For an SIS, conservation of energy gives:
\begin{equation}
v^2=v_p^2+4\sigma^2\ln(r_p/r),
\label{eq:SISfoc1}
\end{equation}
where $v_p$ is the star's velocity at pericenter. If $b$ is the impact
parameter at distance $r$, angular momentum conservation yields 
\begin{equation}
b^2=r_p^2 [1+4(\sigma^2/v^2)\ln(r/r_p)].
\label{eq:SISfoc2}
\end{equation}
The second integral on the right-hand-side of equation (\ref{lossconeint}) can 
then be rewritten as
\begin{equation}
\begin{split}
\int_a^{R_B}& 4\pi r^2\,\rho(r)\, \Theta(r)\,\\
   &\left\{ \int_0^\infty 4\pi\,v^2\, f(v)\, [1+4(\sigma^2/v^2)\ln (r/a)]dv \right\} dr,
\end{split}
\label{eq:SISfoc3}
\end{equation}
where $f(v)$ is the stellar velocity distribution. For a Maxwellian, the above 
equation can be simplified by setting $v=\langle v \rangle=\sqrt{3}\sigma$ in equation 
(\ref{eq:SISfoc2}): one can then define a $\Theta$-factor that includes gravitational 
focusing
\begin{equation}\label{theta2}
\Theta(r)\rightarrow \Theta(r)\simeq 3.1\left[1-\sqrt{1-\left(\frac{a}{r}\right)^2}\right]\,
\left[1+\frac{4}{3}\ln \left(\frac{r}{a}\right)\right].
\end{equation}
Numerical integration of equation (\ref{lossconeint}) finally yields for the stellar mass
in the loss cone
\begin{equation}\label{lossconeSIS2}
M_*\simeq {8.2\sigma^2\over G}\,a\simeq 2\left(\frac{a}{a_h}\right)\,M_2.
\end{equation}
Note that a fraction $\sim 0.5M_2$ of the mass contained in the loss cone when the 
binary becomes hard ($a=a_h$, $M_*\simeq 2\,M_2$) lies within $a_h$.
Let $t=0$ be the time at which the binary separation is $a=a_h$. 
The number flux of stars into the geometrical loss cone, i.e. the flux of stars with 
$r>a_h$ at $t=0$ that interact with the binary at a later time $t$, is
\begin{equation}\label{lossconetime}
{\cal F}\simeq \frac{2\sqrt{3}\sigma^3}{Gm_*\pi a_h^2}\,\Theta,
\end{equation} 
where $\Theta=\Theta(\sqrt{3}\sigma\,t+a_h)$.

\subsubsection{Energy exchange}

Let us denote with ${\cal E}$ the total binding energy of the MBHB, ${\cal 
E}=GM_1M_2/2a$. The total energy transfer rate from the binary to stars in the loss 
cone can be written as 

\begin{equation}\label{energybudget}
\frac{d{\cal E}}{dt} \simeq \frac{\Delta E_*(a) M_*(a)}{t_{\rm ch}},
\end{equation}
where $t_{\rm ch}$ is a characteristic interaction timescale and 
$\Delta E_*(a)\sim G\mu/a$ is the characteristic specific energy gain of stars as 
a consequence of the gravitational slingshot. MM03 have written equation 
(\ref{energybudget}) in the case of {\it returning stars}, i.e. kicked stars 
that do not escape the 
host bulge and can have a secondary super-elastic interaction with the MBHB.
Returning stars have energy $E(a) \sim \phi(r_{\rm inf}) + \Delta E_*(a)$, and their 
interaction timescale $t_{\rm ch}$ can be identified with the typical 
radial period of stars in an SIS potential,
\begin{equation}
\begin{split}
t_{\rm ch}\sim P(E)&= P(0)\exp(E/2\sigma^2)\\ 
       &= P(0)(M/M_B)\exp\left[\left(\frac{2C}{1+q}\right)
       \left(\frac{a_h}{a}\right)-1\right].
       \label{periodSIS}
       \end{split}
\end{equation}
Here, $P(0)=\sqrt{\pi}GM_B/2\sigma^3$ is of order the bulge crossing time, 
and the average dimensionless energy change $C$ is of order unity and nearly 
independent of $a$ for $a<a_h$. As noted by MM03, in this case 
$M_*(a)\,\Delta E_*(a)\propto a^1a^{-1}\sim $const.     
From equation (\ref{lossconeSIS2}) simple calculations lead to
\begin{equation}\label{arereturning}
\frac{a_h}{a}\simeq \frac{a_h}{a_1}+\frac{1+q}{2C}\,\ln\left[1+8\,C^2\,\frac{q}{(1+q)^2}\frac{(t-t_1)}{P_1}\right],
\end{equation}
where $a_1$ is the binary separation at time $t=t_1$ when secondary slingshots start, 
and $P_1$ is the period of stars with energy $E(a_1)$.

MM03's analysis can be expanded to account for the effect of three-body scatterings
when the MBHB first becomes hard at separation $a=a_h$. Stars with $r<a_h$ at $t=0$ will 
interact with the binary within a timescale $t_{\rm ch} \sim a_h/(\sqrt{3}\sigma)$.
Substitution of $t_{\rm ch}$ in equation (\ref{energybudget}), followed by simple 
algebra, leads to the expression 
\begin{equation}\label{afirstin}
\frac{a_h}{a}=1+C\left(\frac{q}{1+q}\right)\left(\frac{\sqrt{3}\sigma}{a_h}\right)t.
\end{equation} 
A further contribution to the shrinking is associated with stars having $r>a_h$ 
at $t=0$ that are bound to enter the loss cone at later times. This population has total 
mass $\sim 1.5\,M_2$, and its contribution to the orbital decay is given by
\begin{equation}\label{enfirstout}       
\frac{d{\cal E}}{dt} \simeq \Delta E_*(a) {m_* {\cal F} \pi a^2}.
\end{equation}
A straightforward substitution gives
\begin{equation}\label{afirstout}
\frac{d}{dt}\left(\frac{a_h}{a}\right)=C\left(\frac{q}{1+q}\right)\left(
\frac{\sqrt{3}\sigma}{a}\right)\,\Theta(\sqrt{3}\sigma\,t+a_h). 
\end{equation}
The above equation holds for a bulge crossing time, $t<R_B/\sqrt{3}\sigma$, and 
must be solved numerically. 

\begin{figure}
\epsscale{1.0}
\plotone{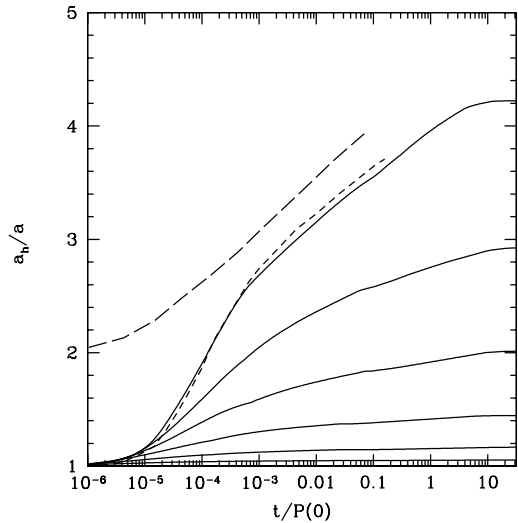}
\caption{Decay of binary separation $a$ (in units of $a_h$) as a function of time 
[in units of the bulge crossing time $P(0)=1.32\times 10^7\,\, {\rm yr}\,\,M_6^{1/4}$ 
for a SIS]. The curves, 
from bottom to top, are for $q=1/243, 1/81, 1/27, 1/9, 1/3, 1$. 
The $q=1$ case is compared to the analytical 
estimate ({\it long-dashed line}) of MM03, and to an N-body simulation ({\it short-dashed line}) of MM03 
performed with 18,000 stars initially in the loss cone, and the stellar potential replaced by a smooth component to prevent 
relaxation. We set $M=0.0014\, M_B$ and use the 
$M-\sigma$ relation (eq.~\ref{msigma}).}
\label{fig:ah-over-a}
\end{figure}

\subsubsection{Orbital decay}

The simple analytical formulation described above can be refined using results from our
scattering experiments (Paper I). This allows us to follow at the same time both the 
orbital decay of the binary and the evolution of the distribution function of interacting 
stars. The procedure is the following. We first isolate, from the initial distribution of 
kicked stars, the {\it new loss cone}, i.e. the subset of stars returning to the center on 
orbits with  $r_p<a_1$, where $a_1$ is the binary separation at the end of the first 
interaction with the stellar background. Then we compute the hardening rate $H$ by averaging 
$H_1(v)$ (provided by our scattering experiments) over the velocity distribution 
function of such stars, which are allowed to interact with the binary for the timescale in 
equation (\ref{periodSIS}), again averaged over the stellar velocity distribution. 
After each step, the velocity distribution function of returning stars is updated, and so 
is the timescale of the following interaction. We iterate the process until the loss 
cone is emptied. Convergence to the final stalling separation is usually obtained after 
$\gtrsim 4$ iterations. The mathematical details of the numerical procedure are 
given in the Appendix.
 
In order to specify the two parameters defining the SIS, the Haring \& Rix (2004) 
bulge-black hole mass relation was complemented by the $M-\sigma$ relation (Ferrarese \& Merrit 
2000; Gebhardt et al. 2000) proposed by Tremaine et al. 2002:
\begin{equation}\label{msigma}
\sigma_{70}=0.84\,M_6^{1/4},
\end{equation}
where $\sigma_{70}$ is the stellar velocity dispersion in units of 70 km/s, and $M_6$ is 
the MBHB mass in units of $10^6\,\msun$. Using equation (\ref{msigma}), the hardening 
radius can then be written as 
\begin{equation}\label{hardradsigma}
a_h\simeq 0.32\,\,{\rm pc}\,\,M_{2,6}^{1/2}\sqrt{\frac{q}{1+q}},
\end{equation}
where $M_{2,6}\equiv M_2/10^6\msun$. The binary separation as a function of time is shown 
in Figure~\ref{fig:ah-over-a}, where it is also compared to the results of an N-body simulation 
and to an analytical prescription, both presented in MM03 (their Fig. 6). The agreement with 
the simulation is fairly good. Our results, in terms of the final separation achieved 
as a function of $q$,  are perfectly consistent with the stalling radii estimated by 
Merritt (2006).
 
Figure~\ref{fig:ah-over-a} shows how the rate of orbital decay declines after a few 
bulge crossing times $\sim P(0)=1.32\times 10^7\,\, {\rm yr}\,\,M_6^{1/4}$:  
this is due to the decreasing supply of low angular momentum stars 
from the outer regions of the bulge, once the stars in the central cusp 
have interacted with the binary. Note that equal mass binaries shrink more than unequal 
binaries: this is both because of the scaling of the stellar mass available for 
the interaction and of the energy exchanged during a typical three-body encounter. 
It is easy to see that $d(1/a)/d\ln t \propto q$. Eccentricity plays a marginal role in the 
orbital evolution of the MBHB. For a given $q$, the orbital shrinking is larger by at 
most $10$\% for highly eccentric binaries. 

\begin{table}
\begin{center}
\caption{Binary hardening: the impact of returning stars }
\begin{tabular}{ccc}
\tableline\tableline
$q$  & $a_h/a_1$ & $a_h/a_f$\\
\tableline
   1&       2.81&     4.41\\
   1/3&     2.19&     3.07\\
   1/9&     1.64&     2.09\\
   1/27&    1.29&     1.49\\
   1/81&    1.12&     1.19\\
   1/243&   1.04&     1.06\\
\tableline
\end{tabular}
\tablecomments{Binary shrinking factors. $a_h/a_1$ is the binary shrinking 
after the first interaction only, while $a_h/a_f$ take into account for
subsequent reejections up to the fourth interaction.}
\label{Tab1}
\end{center}
\end{table}

\begin{figure}
\epsscale{1.0}
\plotone{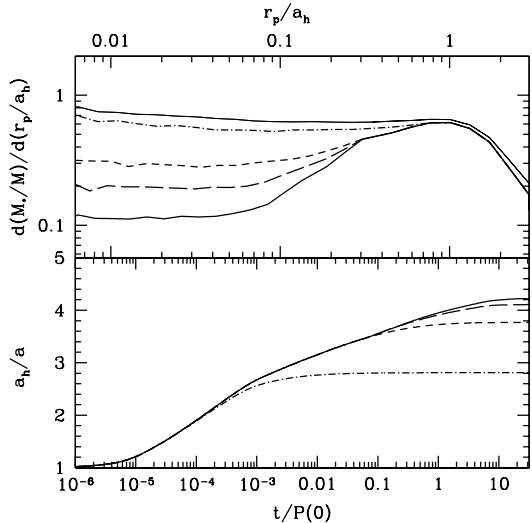}
\caption{{\it Upper panel}: evolution of the loss-cone population in terms 
of the (differential) stellar mass that approaches the binary with a given periastron $r_p$.
{\it Thin line}: initial loss cone.
{\it Solid lines, from top to bottom:} loss-cone population after
the 1st, 2nd, 3rd, and 4th interaction. 
{\it Lower panel}: binary separation as a function of time. From bottom to top, the curves 
depict the shrinking associated with only the first one, two, three, and four interactions, 
respectively. An equal mass, circular binary is assumed.}
\label{fig:returning}
\end{figure}

Figure~\ref{fig:returning} shows an example of the role of secondary slingshots on 
orbital shrinking for an equal mass circular binary. The lower panel clearly illustrates 
how successive interactions of stars returning on quasi radial orbits can reduce the final 
binary separation by an extra factor of order 2, i.e. $a_f \sim a_1/2$.   
The progressive emptying of the loss cone is sketched in the upper panel, where we plot 
the (differential) mass in stars approaching the binary with a given periastron. 
After the first interaction only few stars are kicked out from the bulge. The loss cone, 
while substantially hotter, remains nearly full and only gets progressively depleted as 
secondary slingshots take place.
Table \ref{Tab1} quantifies the role of returning stars for different values 
of the binary mass ratio $q$: returning stars can increase the shrinking of the MBHB 
by as much as a factor of 2, and play a larger role for equal mass binaries. This is 
because binary-to-star energy exchange is significant only for $V\lesssim \sqrt{q} 
V_c$ (see Paper I). After the first binary-star interaction, the stellar population is heated up
and stars have on average $V \sim \sqrt{q}\,V_c$. In the case $q=1$, the binary shrinks by a 
significant factor, $V_c$ increases, and most of the returning stars have $V\lta V_c$: the 
hardening process is still efficient. By contrast, when $q\ll 1$, $V_c$ does not increase 
appreciably after the first interaction, returning stars have $V\sim \sqrt{q} V_c$, and 
binary hardening stops.

\section{Discussion}

\begin{figure}
\epsscale{1.0}
\plotone{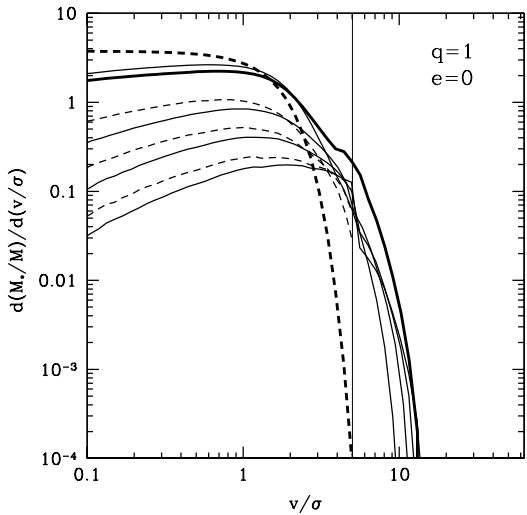}
\caption{Stellar velocity distribution for an equal mass, circular binary, at 
different stages of binary hardening. 
The vertical lines mark $v_{\rm ret}= 5.1 \sigma$ (we recall that $v_{\rm esc}=5.5 \sigma$).
{\it Dashed lines:} from top to bottom, distribution of stars in the shrinking  
loss cone before the 1st, 2nd, 3rd and 4th iteration. For clarity, the initial loss cone distribution is 
marked with a thicker line. {\it Thin solid lines:} from top to bottom, distribution of stars 
that have 
received 1, 2, 3 and 4 kicks. 
{\it Thick solid line:} final stellar velocity distribution after loss-cone depletion is completed.}    
\label{fig:dMdv}
\end{figure}

Under the assumed criterion for stellar ejection, we can compute $M_{\rm ej}$, the mass 
of stars expelled with $V>v_{\rm esc}$.  
An example of the effects of the slingshot mechanism on the stellar population is shown 
for an equal mass circular binary in Figure~\ref{fig:dMdv},
where the initial ($t=0$) velocity distribution of interacting stars 
is compared to the distribution after loss-cone depletion.
As already mentioned in the previous section, after the first interaction
with the binary a large subset of kicked stars still lies in the loss cone of the shrinking
binary, has velocities $v<v_{\rm ret}$, and is potentially avaliable for
further interactions. These are the stars we termed ``returning''.  
While scattering with the binary increases the stellar velocity thus reducing the energy 
exchanged in secondary interactions, it moves kicked stars on more radial orbits thus 
reducing their impact parameter as well.
Our calculations show that the high velocity tail of the distribution depends on the 
MBHB eccentricity $e$, although the effect of changing $e$ is small for small values of $q$. 
In this case, fewer stars are kicked out compared to the case $q\lesssim 1$, but at higher 
velocities on the average. In general, both a small mass ratio and a high eccentricity 
increase the tail of HVSs. Integrating the curves in Figure~\ref{fig:dMdv} over velocity 
gives the mass of interacting stars: this is $\simeq 2M$ for the case shown ($q=1$), 
$\simeq 1.2M$ for $q=1/3$, and $\simeq 0.6M$ for $q=1/27$ (all assuming $e=0$; we checked that 
eccentricity plays a negligible role).

\begin{figure}
\epsscale{1.0}
\plotone{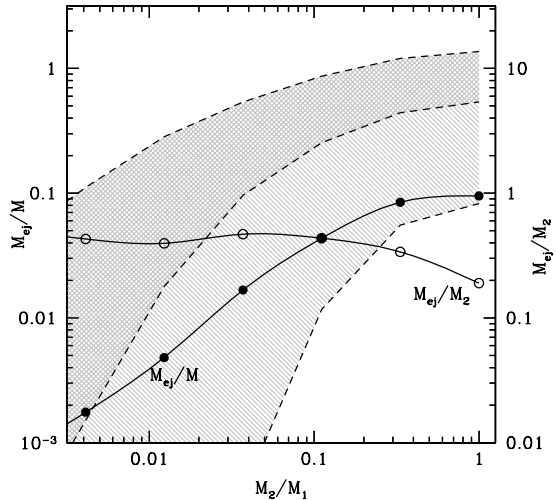}
\caption{Ejected stellar mass $M_{\rm ej}$ normalized to the total binary mass $M$ 
({\it left scale, solid points}), and to the mass of the lighter binary member $M_2$ 
({\it right scale, empty points}), as a function of binary mass ratio. 
The curves are polynomial interpolations. Note that the ratios $M_{\rm ej}/M$ and 
$M_{\rm ej}/M_2$ do not depend on the absolute value of $M$, and are nearly independent on $e$.
{\it Upper, dark shaded area:} the mass (normalized to $M$) 
a $M=10^6 \msun$ binary needs to eject to reach a final separation 
$a_f$ such that $t_{\rm GW}=1$ Gyr. Top and bottom boundaries assume $e=0$ and $e=0.9$, 
respectively. {\it Lower, light shaded area}: same but for a $M=10^9 \msun$ binary. 
}
\label{fig:M_ej}
\end{figure}

\begin{figure}
\epsscale{1.0}
\plotone{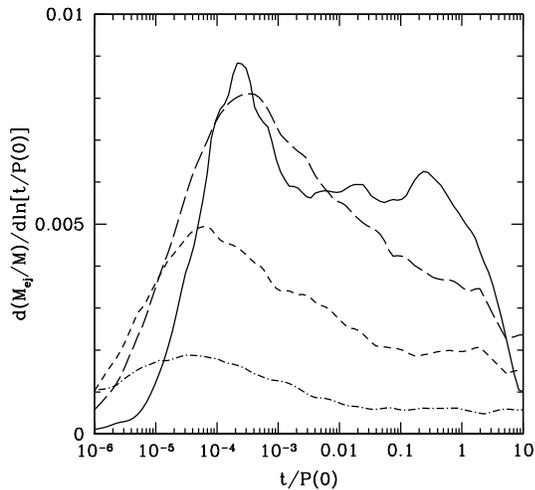}
\caption{Ejected stellar mass per unit logarithmic time interval, as a function of 
time. A circular binary is assumed. 
{\it Solid line:} $q=1$. {\it Long-dashed line:} $q=1/3$. {\it Short-dashed line:} $q=1/9$.  
{\it Dot-dashed line:} $q=1/27$.}
\label{fig:dMtime}
\end{figure}

Figure~\ref{fig:M_ej} depicts the ejected mass $M_{\rm ej}$ normalized 
to $M$ (left scale) and to $M_2$ (right scale), as a function of $q$. Our results 
show that $M_{\rm ej}/M\sim 0.5 \mu/M=0.5 q/(1+q)^2$, i.e., $M_{\rm ej}/M_2 \sim 0.5/(1+q)$,
both ratios being independent of the total binary mass. 
The rate of stellar mass ejection is shown in Figure~\ref{fig:dMtime} as a function of time.
A fraction $\lesssim 50\%$ of the expelled stars 
is ejected in a initial burst lasting $\sim a_h/\sigma$, this fraction being smaller
for smaller binary mass ratios. The burst is associated to the ejection of those stars 
already present within the geometrical loss cone when the binary first becomes hard. 
Note that, for small $q$, mass ejection is already significant at $a\simeq a_h$, as in 
this case the binary orbital velocity is $V_c\gtrsim v_{\rm esc}$ for $a \lesssim a_h$.  

Is the amount of ejected mass sufficient to shrink the MBHB orbit down to the GW-dominated 
regime? To answer this question, we start defining the ``final separation" $a_f$ as 
the separation reached by the binary before complete loss-cone depletion, i.e. 
after a few bulge crossing time ($\sim 10^7$ yrs, weakly depending on binary mass 
as $\propto M^{1/4}$). We must then compare $a_f$ to the separation at which the 
orbital decay timescale from GW emission,
\begin{equation}
\begin{split}
t_{\rm GW}  &=\frac{5c^5}{256 G^3}\frac{a^4}{M_1 M_2 M F(e)}\\
& \approx 0.25\,{\rm Gyr}\left(\frac{M M_1M_2}{10^{18.3}\,\msun^3
}\right)^{-1}\,F(e)^{-1}\,\left(\frac{a}{0.001{\rm pc}}\right)^4
\end{split}
\label{eq:tgw}
\end{equation}
(Peters 1964), is shorter than, say, 1 Gyr. Here, to 4th order in $e$,
\begin{equation} 
F(e)=(1-e^2)^{-7/2}\left(1+\frac{73}{24}e^2+\frac{37}{96}e^4\right).
\label{eq:ecce}
\end{equation}
Inverting equation (\ref{eq:tgw}), one can define the separation 
$a_{\rm GW}$ at which the binary will coalesce in a given time $t$,
\begin{equation}
\begin{split}
a_{\rm GW}  &=\left[\frac{256 G^3}{5c^5}t M_1 M_2 M F(e)\right]^{1/4}\\
& \approx 0.0014\,{\rm pc}\,\left(\frac{M M_1M_2}{10^{18.3}\,\msun^3
}\right)^{1/4}\,F(e)^{1/4}\,t_9^{1/4},
\end{split}
\label{eq:agw}
\end{equation}
where $t_9\equiv t/1\,{\rm Gyr}$. 
Using equations (\ref{hardradsigma}) and (\ref{eq:agw}), one finds that   
$a_f/a_{\rm GW}\propto M^{-1/4}q^{3/4}$ (see also MM03, eq.~90), 
i.e. the more massive the binary and the 
smaller the binary mass ratio, the smaller the factor the binary must shrink 
to reach the GW emission regime. Eccentricity plays a double role. For a given binary mass 
and mass ratio, on one hand the hardening rate slightly increases with increasing eccentricity 
($\lesssim 20$\% from $e=0$ to $e=0.9$), leading to a smaller $a_f$; on the other hand, 
from equations (\ref{eq:ecce}) and (\ref{eq:agw}), $a_{\rm GW}$ is larger for larger $e$ thus 
reducing the $a_h$-$a_{\rm GW}$ gap. An important effect, included in our calculations, 
is that the eccentricity typically increases during the binary-star interaction, though 
the functional form of $F(e)$ is such that the effect is significant only for binaries with 
$e\gtrsim 0.6$ already at $a_h$. In other words, for MBHBs with initially low eccentricities,
the increase of $e$ during the gravitational slingshot affects only weakly the final 
$a_f/a_{\rm GW}$ ratio.

It is interesting to compare the total mass actually ejected prior to complete loss-cone depletion  
with the stellar mass that must be expelled in order to reach 
a final orbital separation where $t_{\rm GW}(a_f)=1$ Gyr, i.e. where
GW emission leads to coalescence within 1 Gyr. An example is shown in Figure~\ref{fig:M_ej}. 
The shaded areas define such mass (in units of $M$) for a $10^6\,\msun$ and a 
$10^9\,\msun$ MBHB, where the top boundary assumes $e=0$ and the bottom $e=0.9$. 
Note how the $e=0.9, M=10^6\,\msun$ lower boundary practically coincides with the $e=0, 
M=10^9\,\msun$ upper one. The figure clearly shows how, even in the absence of other 
mechanism driving orbital decay, pairs involving genuinely supermassive holes should not 
stall, while for lighter binaries both a small mass ratio and a large eccentricity are 
probably required for coalescence to take place. 

\begin{figure}
\epsscale{1.0}
\plotone{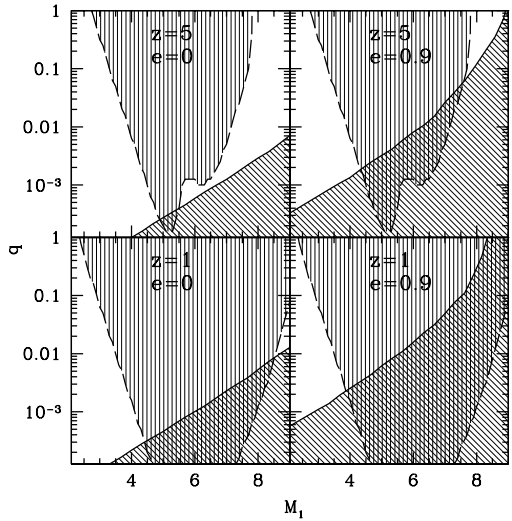}
\caption{$M_1-q$ plane. The vertical shaded area shows {\it LISA} potential targets 
with $S/N>5$. The diagonal shaded area on the lower right corner marks binaries that 
will coalesce within a then Hubble time after loss-cone depletion. In each panel, 
the assumed redshift and eccentricity of the MBHBs are labeled.}
\label{fig:LISA}
\end{figure}

Using our hybrid model, we can also sample the $(M_1,q,e)$ 3-D space, compute the separation 
$a_f$ and the eccentricity $e$ at $a_f$, then fold the calculated values of $a_f$ and $e$ 
into equation (\ref{eq:tgw}), and finally compare $t_{\rm GW}$ to the Hubble time at two 
reference redshifts, $z=1$ and $z=5$. In Figure~\ref{fig:LISA}, binaries that will 
coalesce within a then Hubble time after loss-cone depletion populate the diagonally-shaded 
area in the $M_1-q$ plane, while the vertically-shaded area marks MBHBs that, if driven to 
coalescence by $z=1$ or $z=5$, would be resolved by {\it LISA} with a signal-to-noise ratio 
$S/N>5$ (see Sesana \etal 2005 and references therein for details). The region of overlap 
selects unequal mass, highly eccentric MBHBs with $M\gta 10^5$ that can shrink down to 
the GW emission regime in less than an Hubble time, and that  
are ``safe" targets for {\it LISA} even in the pessimistic case, treated 
here, of stellar slingshots$+$loss-cone depletion with no refilling. 

It is important to remark, at this stage, that our calculations are meant to define a 
minimal model for the evolution of MBHBs, and that several other mechanisms may help
the orbital decay and widen the range of potential {\it LISA} targets.
First, we have assumed all the stars in the loss cone to be unbound to the MBHs. 
In a realistic case, each MBH will bind stars inside its radius of influence 
$r_{\rm inf}$: the star binding energy can be extracted by slingshot, hence 
enhancing binary hardening. This effect is not expected to be important for 
equal mass binaries as, in this case, $a_h\sim r_{\rm inf}$, and only a small 
fraction of interacting stars will be bound to the binary. Indeed, our results match well 
the numerical simulations of MM03 (Fig.~\ref{fig:ah-over-a}). For lower mass ratios, 
however, it is $a_h \ll r_{\rm inf}$ and most stars in the loss-cone are 
actually bound to the binary. A forthcoming paper will be devoted to an analysis of 
three-body scattering experiments for a MBHB 
with bound stars, providing a more realistic model for the case $q\ll 1$. 

Second, even in spherical stellar bulges, loss-cone refilling due to two-body relaxation 
(Yu 2002; MM03) and the wandering of the black hole pair in the nucleus (Quinlan \& 
Hernquist 1997; Chatterjee et al. 2003) could both increase the amount of stellar 
mass interacting with the MBHB. The two-body relaxation timescale is such that loss-cone 
refilling is probably important for $M\lesssim 10^6 \msun$. The Brownian motion timescale 
is of the order of 15 Gyr for a 10$^6 \msun$ binary, and scales as $M^{5/4}$. It is then likely 
that the two effects considered here may affect orbital decay only for light binaries, 
helping them to cover the residual gap between $a_f$ and $a_{\rm GW}$ and leading 
light binaries to coalesce within a then Hubble time even at high redshifts. On the other hand, 
their contribution to the shrinking of supermassive binaries with $M\gtrsim 10^6 \msun$ is probably 
negligible.

If the stellar bulge is not spherical, but axisymmetric, stars on highly eccentric 
orbits are typically centrophilic (Touma \& Tremaine 1997; Magorrian \& Tremaine 1999). 
In this case, the loss cone is substituted by a ``loss wedge'' (see Yu 2002 for a 
detailed discussion). The stellar content of such wedge is larger than that of the 
corresponding loss cone, and depends on the degree of flattening $\epsilon$ 
of the stellar distribution. Typically, for a galaxy with $\epsilon=0.3$, the stellar 
content of the loss wedge is a order of magnitude larger than the stellar content 
of the loss cone were the bulge spherical (Yu 2002). Note that Faber et al. (1997) 
estimate an average $\epsilon=0.36$ for a sample of galaxies, leading to the conclusion 
that the hardening of a MBHB in such a potential might be much faster than our 
``spherical" estimate. We also recall that triaxial potentials drive many bulge stars 
on chaotic orbits, many of them centrophilic: one then expects an increase in the 
number of interacting stars similar to that produced by axisymmetric potentials
(Merritt \& Poon 2004; Berczik \etal 2006).

Finally, MBHB orbital evolution can also, at least partially, be driven by drag in a 
gaseous nuclear disk. The role of gas is, basically, twofold. On 100 pc scales, the 
disk drastically increases dynamical friction, reducing the timescale on which 
MBHs can reach the center of the bulge (Escala et al. 2004; Dotti et al. 2006). On 
parsec scales, torques induced by the disk can drive the binary to decay on a timescale 
of order the gas accretion time (Ivanov et al. 1999; Armitage \& Natarajan 2002). 
It is important to point out that the interaction with a gaseous disk typically
circularizes the binary orbit (Dotti et al.2006), hence maximizing the $a_h-a_{\rm GW}$ gap. If this is the 
case, the slingshot driven coalescence would be more difficult to achieve. 

%

\acknowledgments
\noindent
Support to this work was provided by the Italian MIUR grant PRIN 2004 (A.S. and F.H.), 
and by NASA grants NAG5-11513 and NNG04GK85G (P.M.). 
P.M. also acknowledges support 
from the Alexander von Humboldt Foundation. 

\appendix
\section{Numerical integration of the MBHB orbital decay}

The average hardening rate for a Maxwellian stellar velocity distribution
$f(v,\sigma)=(2\pi\sigma^2)^{-3/2}$ $\exp(-v^2/2\sigma^2)$ is given by
\begin{equation}\label{hsigma}
H(\sigma) \equiv \int_0^{\infty} f(v,\sigma)\frac{\sigma}{v}H_1(v)\,4\pi v^2\,dv,
\end{equation}
where
\begin{equation}\label{H1}
H_1(v) \equiv 8\pi\int_0^{\infty} \langle C\rangle x\,dx
\end{equation}
is the dimensionless hardening rate if all stars have the same velocity $v$, 
$x\equiv b/\sqrt{2GMa/v^2}$ is the dimensionless impact parameter, and 
the energy exchange $\langle C\rangle$ is averaged over the orbital angular variables 
(Paper I; Q96). 
An expression analogous to equation (\ref{hsigma}) relates the thermally-averaged
eccentricity growth rate $K(\sigma)$ to $K_1(v)$:
\begin{equation}\label{K1}
K_1(v) \equiv \frac{(1-e^2)}{2e}\frac{\int_0^{\infty}\langle B-C\rangle x\,dx}
{\int_0^{\infty}\langle C\rangle x\,dx},
\end{equation}
where $\langle B-C\rangle$ is the mean angular momentum minus energy exchange. 
For a binary with given mass ratio
and eccentricity, the quantities $C$ and $B$, and thus $H_1$ and $K_1$, are only 
function of the ratio 
$v/V_c\propto v\sqrt{a}$, where $V_c$ is the binary circular velocity. 
Given an incoming velocity $v$, we record the bivariate distribution $h_1(V,b'|v)$ 
of stars with ejection speeds in the interval $V,\,V+dV$, leaving the binary with 
an ``exit" impact parameter 
in the interval $b',\,b'+db'$. The distribution function is normalized so that
\begin{equation}\label{eq:fnorm}
\int_0^\infty\int_0^\infty \,h_1(V,b'|v)\,dV\,db'\,=1.
\end{equation}
The subscript ``1'' indicates that the scattering experiments are performed 
for a binary at separation $a=1$.

The interaction of the MBHB with stars in the loss cone is assumed to take place
in discrete steps. The binary first interacts with a given population of stars 
and shrinks accordingly. We then isolate the returning sub-population, which becomes 
the input for the next step, and so on. Consider the binary at separation $a_i$, 
interacting  with a stellar population of mass $M_{*,i}$ and (normalized) velocity 
distribution $f_i(v)$. The orbit decays according to the differential equation
\begin{equation}\label{eq:sepevol}
\frac{d}{dt}\left(\frac{1}{a}\right)\,=\,\frac{G\rho}{<v>}\,H(a),
\end{equation}
where
\begin{equation}\label{eq:Have}
H(a)\,=\,\int_0^{\infty}\,f_i(v)\frac{<v>}{v}\,H_1(v\sqrt{a})\,dv,
\end{equation}\label{eq;aint}
and $\rho$ is the stellar density. Straightforward integration of equation (\ref{eq:sepevol}) gives
\begin{equation}\label{eq:aint}
t(a)\,=\,\frac{<v>}{G\rho}\,\int_{a}^{a_i}\,\frac{da'}{a'^2H(a')}, 
\end{equation}
the time the orbit needs to shrink from $a_i$ to $a$.
The solution above is not physically meaningful, as the time variable involved 
depends on the particular value assumed for 
$\rho$. 
Stated more directly, to solve equation (\ref{eq:sepevol}) we need to set 
a realistic pace at which star-binary interactions occurs. This can be done in two steps. 

First, we write the stellar 
mass $M_*$ that will interact with the binary in a given time $t$ as
\begin{equation}\label{eq:dmdt}
M_*(t)=k_*\,t, 
\end{equation}
where
\begin{equation}
k_*\,\equiv\,\int_0^{\infty}\,f_i(v)\,\pi b_{\rm max}^2(v)\,\rho\,v\,dv,
\end{equation}
and the maximum allowed impact parameter is
\begin{equation}\label{eq:bmax}
b_{\rm max}^2(v)=a_i^2\left(1+\frac{2GM}{a_i v^2}\right).
\end{equation}
Equation (\ref{eq:dmdt}) is valid as long as $M_*(t)\leq M_{*,i}$. 
Now we simply write $t=M_*/k_*$, and substitute into equation (\ref{eq:aint}), which now gives 
$M_*(a)$, the stellar mass interacting with the binary as the orbit shrinks from $a_i$ to $a$. 
Note that in $M_*(a)$ term $\rho$ cancels out, i.e., the interacting mass is independent of any 
pre--assigned value for the stellar density. 
We can now numerically invert the equation for $M_*(a)$, obtaining $a(M_*)$, i.e., the binary 
separation as a function of the interacting mass. We define the final separation $a_f\equiv a(M_{*,i})$. 
The very same procedure can be applied for the evolution of the binary eccentricity $e$, 
resulting in a function describing $e$ as a function of the interacting mass, $e=e(M_*)$. 

We need now to relate $M_*$ to physical time. 
The stellar mass interacting with the binary per unit time is
\begin{equation}\label{eq:dmdt2}
\frac{dM_*}{dt}=\frac{dM_*}{dv}\frac{dv}{dt}\,=\,f_i(v)M_{*,i}\,\frac{dv}{dt},
\end{equation}
where the term $dv/dt$ can be computed considering the typical interaction time for 
stars with velocity $v$ in a SIS potential, 
i.e., $t(v)=P(0)\exp{[(v^2-v_{\rm ret}^2)/4\sigma^2]}$. Straightforward algebra 
yields $dv/dt$ as a function of $t$. 
Equation (\ref{eq:dmdt2}) can be then integrated, and the resulting $M_*(t)$  
finally substituted into $a(M_*)$ to give the time evolution of the binary separation $a(t)$.

We can now compute the distribution of stars that, after the interaction, 
have velocities $V< v_{\rm ret}$, and are then available for a further encounter with the MBHB. 
The bivariate distribution $h_a(V,b'|v)$ can be extracted from the 
$h_1(V,b'|v)$ distributions recorded in the scattering experiments. In the three-body 
problem integration, the binary mass and separation are taken as unity. 
This set-up allows to rescale the velocities and trajectories of kicked stars for 
any given physical value of the binary mass and separation. It can be shown that
\begin{equation}\label{eq:scatdistr}
h_a(V,b'|v)=\frac{1}{\sqrt{a}}\,\times\,h_1\left(\frac{V}{\sqrt{a}},b'a|v\sqrt{a}\right),
\end{equation}
where the prefactor $1/\sqrt{a}$ normalizes the distribution according to 
equation (\ref{eq:fnorm}). The normalized distribution of scattered stars must be 
averaged over the input velocity $v$ and the separation $a$, and can be written as
 \begin{equation}\label{eq:fvfin}
h(V,b')=\frac{\int_{a_i}^{a_f}\int_0^\infty
f_i(a,v)\,(dM_*/da)\,h_a(V,b'|v)\,dv\,da}
{\int_0^\infty\int_0^\infty\int_{a_i}^{a_f}\int_0^\infty
f_i(a,v)\,(dM_*/da)\,h_a(V,b'|v)\,dv\,da\,dV\,db'}.
\end{equation} 
Here, the distribution of stars with 
velocity between $v$ and $v+dv$ that are going to approach the binary 
within a distance $<a$ is  
\begin{equation}\label{eq:fia}
f_i(v,a)=\frac{f_i(v)\pi b^2(v,a)}{\int f_i(v)\pi b_{\rm max}^2(v)\,dv},
\end{equation}
where
\begin{equation}\label{eq:bee}
b^2(v,a)=a^2\left(1+\frac{2GM}{a v^2}\right),
\end{equation} 
and $b_{\rm max}(v)$ is given by equation (\ref{eq:bmax}).
The change of the interacting mass with binary separation $dM_*/da$ is obtained by 
differentiation of the function $M_*(a)$, obtained above.


Finally, the
(normalized) velocity distribution of returning stars $f_r(v)$ can be computed as
\begin{equation}\label{eq:fret}
f_r(v)=\frac{\int_0^{a_f}h(V,b')\,db'}{\int_0^{v_{\rm{ret}}}\int_0^{a_f}h(V,b')\,db'\,dV},
\end{equation}
and the mass avaiable for the subsequent interaction $M_{*,r}$ is given by
\begin{equation}\label{eq:mert}
M_{*,r}=M_{*,i}\frac{\int_0^{v_{\rm{ret}}}\int_0^{b(v,a_f)}h(V,b')\,db'\,dV}
{\int_0^\infty\int_0^\infty h(V,b')\,db'\,dV}.
\end{equation}
The numerical procedure can be then iterated, considering the binary at a starting separation
$a_i=a_f$, interacting with a stellar population whose (normalized) velocity
distribution is $f_i(v)=f_r(v)$, allowing a total mass of interacting stars $M_{*,i}=M_{*,r}$. 
For the first interaction only, we assume that the stars already within the 
binary separation interact on 
a time scale $\sim a_h/\sigma$, along the lines discussed in Section 2.2.2.

{}

\end{document}